\documentclass[pre,twocolumn,superscriptaddress,floatfix,amssymb,showpacs]{revtex4-1}
\usepackage{enumerate}
\usepackage{times}
\usepackage{graphicx}
\usepackage{amsmath}
\usepackage{color}
\usepackage{natbib}
\usepackage{array}
\newcolumntype{C}[1]{>{\centering\let\newline\\\arraybackslash\hspace{0pt}}m{#1}}

\usepackage[pdfpagemode=None,colorlinks=true,urlcolor=black,%
linkcolor=blue,citecolor=blue,pdfstartview=FitH]{hyperref}
\bibpunct{[}{]}{,}{n}{}{;}

\graphicspath{{fig/}} 
\begin{document}

{
\thispagestyle{empty}
\addtocounter{page}{-1}
\onecolumngrid
\raggedright{}
\Huge 
Copyright Notice\\
\vspace{10mm}
\normalsize
Copyright (2014) by the American Physical Society.\\
\vspace{10mm}
Criteria for minimal model of driven polymer translocation\\
P. M. Suhonen, K. Kaski, and R. P. Linna\\
\vspace{10mm}

Citation: Phys. Rev. E 90, 042702 (2014)\\
URL: \href{https://journals.aps.org/pre/abstract/10.1103/PhysRevE.90.042702}{https://journals.aps.org/pre/abstract/10.1103/PhysRevE.90.042702}\\
DOI: 10.1103/PhysRevE.90.042702\\
\clearpage
}

\title{Criteria for minimal model of driven polymer translocation}

\author{P. M. Suhonen}
\author{K. Kaski}
\author{R. P. Linna}
\email{Author to whom correspondence should be addressed: riku.linna@aalto.fi}
\affiliation{Department of Biomedical Engineering and Computational Science, Aalto University, P.O. Box 12200, FI-00076 Aalto, Finland}


\begin{abstract}
While the characteristics of the driven translocation for asymptotically long polymers are well understood, this is not the case for finite-sized polymers, which are relevant for real-world experiments and simulation studies. Most notably, the behavior of the exponent $\alpha$, which describes the scaling of the translocation time with polymer length, when the driving force $f_p$ in the pore is changed, is under debate. By Langevin dynamics simulations of regular and modified translocation models using the freely-jointed-chain polymer model we find that a previously reported incomplete model, where the {\it trans} side and fluctuations were excluded, gives rise to characteristics that are in stark contradiction with those of the complete model, for which $\alpha$ increases with $f_p$. Our results suggest that contribution due to fluctuations is important. We construct a minimal model where dynamics is completely excluded to show that close alignment with a full translocation model can be achieved. Our findings set very stringent requirements for a minimal model that is supposed to describe the driven polymer translocation correctly. 
\end{abstract}

\maketitle
\section{Introduction}\label{sec:intro}

Driven polymer translocation is a process where a polymer chain is driven through a small pore in a membrane by an electric potential applied across the membrane. It has been an active field of research since the pioneering experimental work by Kasianowicz et al.~\cite{Kasianowicz96}, in which they showed translocation to have a potential application in DNA sequencing. Vast amount of work has been done to make it a viable option to the current sequencing methods, see {\it e.g.}~\cite{Branton08,Wanunu12}. In addition, polymer translocation process is normal cellular activity that takes place for example when proteins are imported into mitochondrial matrix\cite{Alberts94}.

The theoretical understanding of the driven polymer translocation has evolved from the first derivations using close to equilibrium framework~\cite{Sung96,Muthukumar99} to considerations of the polymer remaining only marginally in equilibrium during translocation~\cite{Chuang01,Kantor04}. By simulations we have previously shown that the driven polymer translocation is an out-of-equilibrium process~\cite{Lehtola09,Lehtola08}, where the polymer is continuously driven further out of equilibrium on both sides of the membrane. On the {\it cis} side, from which the polymer translocates, there is a growing region where the polymer is under tension and the monomers are in motion. On the {\it trans} side monomers crowd close to the pore exit. We presented a simple sketch to explain the obtained scaling of the translocation time $\tau$ with the finite polymer length $N$, $\tau \sim N^\alpha$, where $\alpha = 1+\nu-\chi$, and $\nu$ is the Flory exponent, and $\chi$ is a constant. This crude derivation was based on writing down a force balance equation for the drag due to the moving segments on both {\it cis} and {\it trans} sides and the driving force inside the pore $f_p$. The sketch was based on very approximate data on how the tension spread on the {\it cis} side and it did not take into account variations in waiting time during translocation. Asymptotically, {\it i.e.} for zero friction corresponding to $N \to \infty$ and large $f_p$, we noted that $\chi \to 0$, and $\alpha = 1+\nu$.

A detailed analytical treatment was given earlier by Sakaue~\cite{Sakaue07}. Here, it was derived how the tension propagates on the {\it cis} side setting  monomers there in motion. This analytical treatment has been adopted and expanded in~\cite{Dubbeldam12}. Sakaue's concept has been given further confirmation by a generalized computational model~\cite{Ikonen12}. The model has been improved in~\cite{Saito12}. Rowghanian and Grosberg have derived a comprehensive and quite conclusive theory for polymer translocation in the asymptotic limit of very long polymers~\cite{Rowghanian11}. At this asymptotic limit the translocation was confirmed to scale as $\tau \sim N^{1+\nu}$. In all the computational work the polymers are inevitably well below the length required for obtaining asymptotic scaling. A finite-size scaling presented in~\cite{Ikonen13} shows the close connection of zero pore friction and the asymptotic limit.

It is fair to say that the asymptotic characteristics of the driven translocation have been derived and proven. There are, however, important open questions for finite polymer lengths, where all experiments and simulations are performed. One notable issue is how $\alpha$ changes with $f_p$. In our earlier simulational work~\cite{Lehtola09} we have measured $\alpha$ increasing  with $f_p$ both with and without hydrodynamic interactions. We addressed this to be due to the observed crowding of monomers close to the pore opening on the {\it trans} side.  $\alpha$ increasing  with $f_p$ was also obtained in a molecular dynamics simulation~\cite{Dubbeldam12}. In contrast, in the numerical model, where the {\it trans} side was not included, $\alpha$ was seen to decrease with increasing $f_p$~\cite{Ikonen12}.

For finite polymer lengths the {\it trans} side, where the polymer is also driven out of equilibrium, may have a significant effect. In addition, fluctuations have recently been shown to facilitate translocation~\cite{Dubbeldam13}. Also fluctuations were omitted from the numerical model, where $\alpha$ was found to decrease with increasing $f_p$. Here, we estimate the importance of the monomer crowding on the {\it trans} side and the pertinent fluctuations to the driven polymer translocation for finite polymer lengths. Coincidentally, an analytical derivation accompanied by numerical solution was recently conducted to address the role of the crowding~\cite{Dubbeldam14}. We also measure the tension spreading on the {\it cis} side in detail. By constructing a quasi-static model for the driven translocation we asses how well a model incorporating the correct initial conformation but where dynamics is completely excluded describes the driven polymer translocation process. This gives us an idea of the precision required of a model constructed to reproduce the characteristics of the process in detail. Finally, we summarize the necessary ingredients of such a minimal model. 

The paper can be outlined as follows. The complete computational model for the driven polymer translocation is described in Section~\ref{sec:cm}. The modified models, the results obtained via simulating them, and the related analysis are presented in Section~\ref{sec:res}. Summary and conclusions are made in Section~\ref{sec:con}.

\section{The computational model}\label{sec:cm}
\subsection{The polymer model}\label{sec:pm}
A coarse-grained freely-jointed spring-bead polymer model is used in all simulations. The model consists of beads connected together as a chain using finitely extensible nonlinear elastic (FENE) potential. The FENE potential is described as

\begin{align}\label{equ:fene}
U_F = -\frac{K}{2}R^2 \ln{\left(1-\frac{r^2}{R^2}\right)},
\end{align}
where $r$ is the current length of the bond and $R=1.5\sigma$ is the maximum bond length. Excluded volume interactions between all beads are implemented by the shifted and truncated Lennard-Jones (LJ) potential

\begin{align}\label{equ:LJ}
U_{LJ} = 4 \epsilon \left[ \left(\frac{\sigma}{r}\right)^{12} - \left(\frac{\sigma}{r}\right)^{6} + \frac{1}{4}\right], r \leq 2^{1/6}\sigma,
\end{align}
where $r$ is the distance between the beads. The values for the parameters above are chosen to be $K=\frac{30}{\sigma^2}$, $\epsilon=1.0$ and $\sigma=1.0$. Good solvent condition is implemented by applying the LJ potential only where it is repulsive. For $r > r_0 = 2^{1/6}\sigma$, $U_{LJ} = 0$. $r_0$ can be regarded as the bead radius. For clarity, the beads in Figures~\ref{fig:ModelGeometry} and~\ref{fig:2BondLengthIllustration} are depicted much larger than this. The length scale can be related to physical length scale for instance by the relation $b = 2 \lambda_p$, where $b = 1$ is the bond length and $\lambda_p$ is the persistence length, {\it e.g.} $40$ \AA \ for a ssDNA~\cite{Tinland97}.

\subsection{The dynamics}\label{sec:srd}
All the simulations are performed by using Ermak's implementation of Langevin dynamics~\cite{Ermak80}. The Langevin equation can be given as
\begin{align}\label{equ:langevin}
\dot{\textbf{p}}_i=-\xi\textbf{p}_i+\pmb{\eta}_i(t)+\textbf{f}(\textbf{r}_i),
\end{align}
where $\textbf{p}_i$ is the momentum, $\xi$ the friction constant, $\pmb{\eta}_i(t)$ the random force, and $\textbf{f}(\textbf{r}_i)$ the resultant of polymer's intrinsic forces and the external driving force. Here the driving force is applied only on beads inside the pore. 

The integration of the Langevin equation is done by using the velocity Verlet algorithm~\cite{vanGunsteren77}. The parameter values are given in reduced units. We set the Boltzmann constant $k_B = k = 1$. The temperature used in the simulations is $T^* = kT/\epsilon = 1$. Forces in reduced units are $\vec{f}^* = \vec{f}\sigma/\epsilon = \vec{f}$ and the time step $\delta t^* = (\epsilon/{m \sigma^2})^{1/2} \delta t = 0.001$~\cite{Allen}. The value for the friction coefficient is $\xi=0.5$ and $\eta_i(t)$ is related to it according to the fluctuation dissipation theorem. The mass for each polymer bead and the time step for the integration are chosen to be $m=16$. 

\subsection{The pore and membrane}
\begin{figure}[t]
\includegraphics[width=0.90\linewidth]{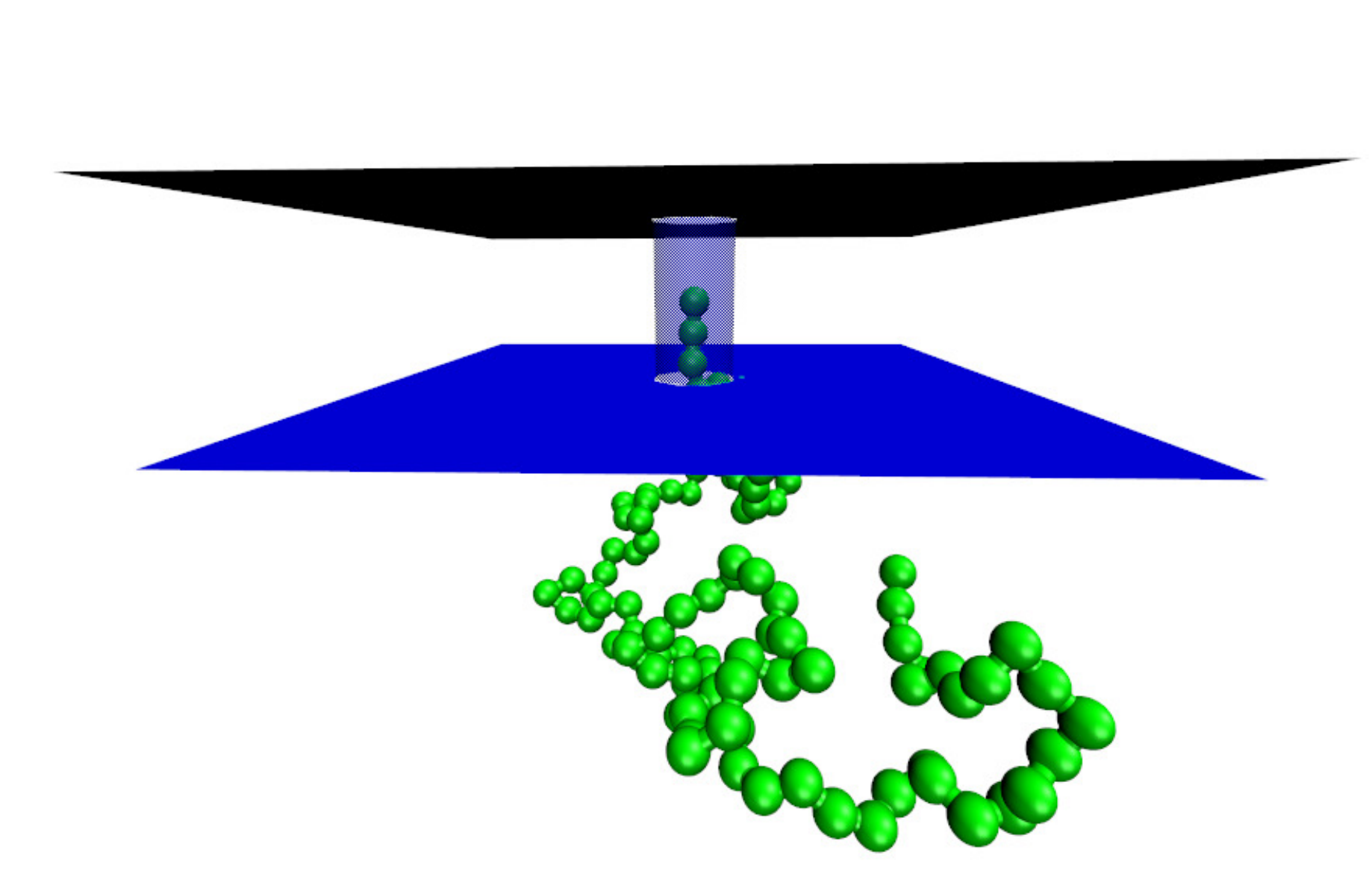}
\caption{(Color online) The simulation geometry, where the polymer is at its initial conformation on the {\it cis} side. Two infinite planes divide the space into {\it cis} and {\it trans} sides. The pore connects these sides allowing the polymer to pass.}
\label{fig:ModelGeometry}
\end{figure}

For the physical model of the membrane we use two aligned infinite planes separating the simulation space into two semi-infinite compartments, as seen in Fig.~\ref{fig:ModelGeometry}. Slip boundary conditions are applied for polymer beads hitting the walls. The separation between the planes in our model is $5\sigma$. For modeling a pore through the membrane, we insert circular holes of diameter $2.25\sigma$ in both planes. Inside the cylindrical pore a linear force pulls the polymer beads toward the center axis according to 
\begin{align}\label{equ:harmonicpore}
f_h= -k_p r_p -c v_p,
\end{align}
where $r_p$ is the distance from the pore axis and $v_p$ the velocity component perpendicular to the pore axis. The parameter values are chosen as $k_p=100$ and $c=1.0$. $f_h$ keeps the polymer straight inside the pore preventing hair pinning. The driving pore force $f_p$ is given as force per bead (or monomer). Hence, the total pore force is $f^{tot}_p = 5 \sigma f_p$. ($f^{tot}_p$ is kept constant in spite of variations in the number beads inside the pore by adjusting momentary force per bead so that the pore exerts a constant force per polymer length.) To estimate {\it e.g.} the magnitude of the total pore force in SI-units, when $f_p = 1$ in reduced units, one writes $\tilde{f} \tilde{b} / k_B \tilde{T} = f_pb/kT = 1$ to obtain $\tilde{f_p} = k_B \tilde{T}/\tilde{b} = 0.52$ pN. Hence, $\tilde{f}_{tot} = 5 \tilde{f_p} = 2.6$ pN. For a typical potential of $120$ mV driving a polymer through an $\alpha$-HL pore and taking the effective charge per nucleotide to be $q \approx 0.1 e$, where $e$ is the elementary charge~\cite{Sauerbudge03,Luan08,Keyser06}, and assuming for a ssDNA roughly ten nucleotides per $40$ \AA\ , we obtain $\tilde{f}_{tot} \approx 5$ pN. Obviously, due to their dependence {\it e.g.} on the pore structure, precise estimates of realistic pore force magnitudes are impossible to make. Nevertheless, the pore force in our simulations is of the right order of magnitude.

\section{Results}\label{sec:res}

Here, we present results obtained from our unmodified and modified polymer translocation models. Our aim is to determine, how completely driven translocation dynamics is determined by the tension spreading on the {\it cis} side and to what extent the {\it trans} side affects the characteristics. We have previously shown that the polymer segment on the {\it trans} side, along with the segment on the {\it cis} side, is continuously driven further out of equilibrium~\cite{Lehtola09,Lehtola08}. We first characterize this for the present model in~\ref{rg} and continue to asses the roles of both sides by modifying the model in~\ref{br}. Then we measure from simulations of the full dynamical model the length of the tensed segment. We construct a quasi-static model void of any dynamics to see how well this crude model describes the process in~\ref{wt}. Unless noted otherwise, all measurements are averages of $500$ simulated translocations. In all figures error bars are smaller than the symbols marking the measured values.

\subsection{Radius of gyration on the \textit{trans} side}
\label{rg}

\begin{figure}[t]
\includegraphics[width=0.90\linewidth]{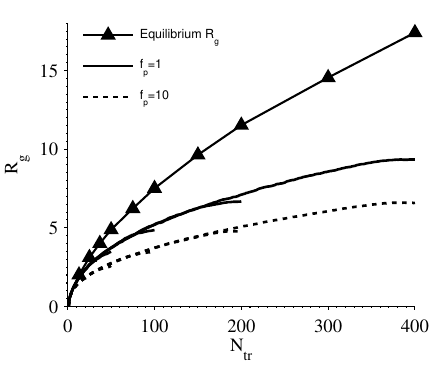}
\caption{Radii of gyration $R_g$ for the translocated polymers of lengths $N = 50$, $100$, $200$, and $400$ for pore force $f_p = 1$ and $10$. (Plots for polymers of different lengths are seen to end at their respective lengths.) The measured $R_g$ on the {\it trans} side are seen to be clearly smaller than the equilibrium $R_g$ for segments of equal lengths.}
\label{fig:TransRG}
\end{figure}

Similarly as for our model including hydrodynamics~\cite{Lehtola09}, we compute the radius of gyration $R_g$ of the polymer segment on the {\it trans} side as a function of the number of beads on the \textit{trans} side $N_{tr}$. This way we can determine how far out of equilibrium the polymer on the \textit{trans} side is driven during translocation. A weak external pore force of $f_p=1$ is applied to all beads within the pore. $R_g(N_{tr})$ for five different chain lengths are given in Fig.~\ref{fig:TransRG}. We compare these to $R_g(N=N_{tr})$ of equilibrated chains attached to a wall of identical boundary conditions. The computed $R_g$ for ten different $N$ are averages over $200$ equilibrated conformations.  $R_g \sim N^\nu$, where $\nu \approx 0.6$ was obtained both for chains attached to the wall and free polymers. Hence, the effect of the wall is negligible. $\nu \approx 0.6$ is the value expected for a polymer with excluded volume interactions in free space.

The plots of $R_g$ in Fig.~\ref{fig:TransRG} as a function of $N_{tr}$ show that the polymer segment on the {\it trans} side is driven increasingly out of equilibrium during translocation. Since the polymer translocates much faster than it relaxes toward thermal equilibrium, the monomers crowd and the polymer conformation is strongly compressed on the {\it trans} side. $R_g$ for polymers of different lengths evolve identically. Only at the very end of the translocation $R_g(N_{tr})$ deviates from the general trend due to the speed up as the end of the chain is pulled through the pore. This speed up can be seen in the waiting time profile, Fig.~\ref{fig:WaitTime}.

So, even for the fairly moderate pore force $f_p = 1$ crowding on the {\it trans} side is considerable and can potentially slow down the translocation. This slowing down would be stronger for long polymers. Increasing $f_p$ naturally enhances crowding, as can be seen in Fig.~\ref{fig:TransRG}, where the \textit{trans} side $R_g$ for polymers of same length are shown also for $f_p=10$. Hence, crowding on the {\it trans} side could cause the $\alpha$ to increase with $f_p$.

\subsection{Roles of the {\it trans} and {\it cis} sides}
\label{br}

Next, we want to asses to what extent the strong crowding on the {\it trans} side affects translocation dynamics, mainly determined by the tension spreading on the {\it cis} side~\cite{Sakaue07,Lehtola09,Lehtola08,Rowghanian11,Dubbeldam12}.

It is practically impossible to determine the effect of crowding during a simulation precisely, since there is no way of isolating it. For example, attempts to measure the force exerted by segments on the {\it trans} side on the monomer at the pore exit are bound to fail due to the difficulty of excluding the force propagated in the chain from the {\it cis} side and the large fluctuations in its magnitude. Due to these difficulties, we assess the effect of the {\it trans} side by simulating modified systems, where either {\it trans} or {\it cis} sides are eliminated. Hence, in these modified systems polymers are either absorbed in or ejected from a wall. Obviously then, the momentum conservation is broken, and the effect caused by this cannot be excluded. In spite of this, conclusions about the roles of the {\it cis} and {\it trans} sides can be drawn.

We compare $\alpha$ for our different model systems. The scaling $\tau \sim N^\alpha$ was measured for different $f_p$. In all model systems the simulation starts with the first bead of the polymer positioned in the middle of the pore, as shown in Fig.~\ref{fig:ModelGeometry}. The polymer is first equilibrated while the first bead is being held fixed. Time required for equilibration was determined by monitoring $R_g$ and seeing that it had stopped diminishing and reached the equilibrium value. The first bead was released after this equilibration time. A polymer bead is considered translocated once it has passed the middle of the pore and the whole process ended when the last bead has translocated.

The modified models were implemented as follows. In the ``no {\it trans}'' model, a polymer bead is removed from the \textit{trans} side when a new bead arrives there. At most two beads are allowed on the \textit{trans} side at any instant. In spite of the out-of-equilibrium character of the process, short distance transitions  were found to dominate driven translocation~\cite{Linna12}. Hence, there is a considerable amount of back-and-forth motion, and by allowing two beads on the {\it trans} side we avoid unnecessary removals and additions of beads. In the ``no {\it cis}'' model new beads are generated at the entrance of the pore as the polymer translocates. The new bead is always placed at equilibrium distance from the bead inside the pore close to the pore opening on the {\it cis} side. In this way no additional force is introduced when generating a bead on the {\it cis} side except for the small extra drag during the short interval in which the generated bead is accelerated. Analogously to the ``no {\it trans}'' model, a buffer of two beads was allowed during backsliding of the polymer.

\begin{figure}[t]
\includegraphics[width=0.90\linewidth]{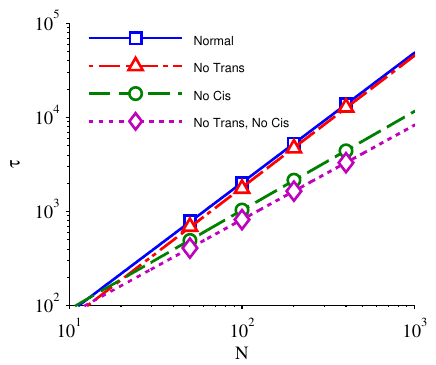}
\caption{(Color online) Translocation times for polymer plotted against polymer length using different kinds of bead removal. From top down: unmodified, no \textit{trans}, and no \textit{cis} translocation models (see text for details). Transfer times for only pore beads are given for reference. Pore force $f_p = 1$.}
\label{fig:ScalingBeadRemoval}
\end{figure}

Translocation times $\tau$ as a function of the polymer length are given for $f_p = 1$ in Fig.~\ref{fig:ScalingBeadRemoval}. The lines are least squares fits to the power law $\tau = c N^{\alpha}$. Scaling exponents obtained for different $f_p$ are listed in Table~\ref{tab:ScalingExponents}. It is seen that the removal of the \textit{trans} side beads has a very small effect on the scaling for $f_p = 1$. In contrast, when the segment on the {\it cis} side is excluded, close to linear dependence of $\tau$ on $N$ is obtained. So, as expected, the dynamics on the \textit{cis} side almost solely determines the translocation dynamics. The {\it trans} side slightly slows down the translocation. We also checked that excluding both sides $\alpha = 1$ is obtained, as it trivially should be. Including only the {\it trans} side segment gives $\alpha$ slightly greater than one, so at least in the limit of zero friction on the {\it cis} side the weak effect of crowding can be seen. 

The polymer segments on both sides are driven further away from equilibrium when increasing $f_p$. As can be seen in the first column of Table~\ref{tab:ScalingExponents}, $\alpha$ increases with increasing $f_p$ confirming our previous finding from simulations using stochastic rotation dynamics~\cite{Lehtola09} and the result from MD simulations~\cite{Dubbeldam12}. In contrast, $\alpha$ decreasing with increasing $f_p$ was obtained using a numerical model, where the effect of the {\it trans} side was ignored~\cite{Ikonen12}. In~\cite{Lehtola09} we addressed this increase of $\alpha$ with $f_p$ to be due to stronger crowding on the \textit{trans} side for large $f_p$, see Fig.~\ref{fig:TransRG}.

In the second column of Table~\ref{tab:ScalingExponents} it can be seen that $\alpha$ increases, albeit more weakly, with $f_p$ also in the ``no \textit{trans}'' model. This means that the the increase of $\alpha$ with $f_p$ is not due to crowding. We have previously shown that fluctuations are significant both in driven translocation~\cite{Linna12} and in the related process of capsid ejection~\cite{Linna14}. The increase of $\alpha$ can be addressed to the contribution due to fluctuations, which were also ignored in the model reported in~\cite{Ikonen12}. Dubbeldam {\it et al.} have recently analyzed the role of fluctuations in driven polymer translocation~\cite{Dubbeldam13}. They found that fluctuations assist translocation. The contribution of fluctuations naturally increases with decreasing $f_p$. Hence, for moderate $f_p$, smaller $\alpha$ results, in other words, $\alpha$ increases with $f_p$.

Our simulations support the finding that fluctuations contribute to the increase of $\alpha$ with $f_p$. From the first two columns of Table~\ref{tab:ScalingExponents} it is seen that excluding the {\it trans} side results in increased $\alpha$ for moderate $f_p$. For large $f_p$, $\alpha$ of the unmodified and ``no {\it trans}'' models are equal. Excluding the {\it trans} side results in fluctuations becoming less prominent. There are two possible explanations for this. Firstly, the fluctuating force from the monomers crowded on the {\it trans} side in the complete model are not mediated to the pore in the ``no {\it trans}'' model. The second mechanism for fluctuations becoming less prominent in the ``no {\it trans}'' model is somewhat less direct. The total force exerted on the monomers inside the pore can be written as $f_{tot} = f_p - f_{count} - f_{frict}$, where $f_{count}$ is the osmotic force due to crowding and $f_{frict}$ is the total frictional force. Excluding the {\it trans} side sets $f_{count} = 0$ thus increasing $f_{tot}$. This, in turn, increases the role of drift compared to diffusion, {\it i.e.}, fluctuations, resulting in the increase in $\alpha$ when excluding the {\it trans} side~\cite{Dubbeldam14}. (It should also be noted that the contributions from tension propagation and tail retraction as described by Dubbeldam et al.~\cite{Dubbeldam12} would give similar characteristics also when the {\it trans} side is excluded.)

Hence, the smaller $\alpha$ in the unmodified model for moderate $f_p$ is in accord with the finding of Dubbeldam et al. The weaker increase of $\alpha$ with $f_p$  would then result from the exclusion of the {\it trans} side, which diminishes the role of fluctuations. The values of $\alpha$ in the third column of Table~\ref{tab:ScalingExponents} would also support this finding: When excluding fluctuations on the {\it cis} side $\alpha$ is slightly larger for moderate $f_p$ and diminishes toward unity only at large $f_p$, where fluctuations are negligible compared to the driving through the pore. 

Setting the pore friction close to zero, $\xi = 0.001$, results in increased $\alpha$, as can be seen in the fourth column of Table~\ref{tab:ScalingExponents}. This is in agreement with our previous findings: $\alpha$ decreases with increasing $\xi$, and for zero friction and large $f_p$, $\alpha = 1 + \nu$ is approached from below~\cite{Lehtola09}. (However, $\alpha$ seems to increase with the friction of the {\it whole} system, the pore included, see~\cite{Lehtola10}.) The finite-size analysis of the pore friction~\cite{Ikonen13} is qualitatively supported by these characteristics. Dubbeldam et al. argued that $\alpha$ remains smaller than the asymptotic value $1 + \nu$ due to fluctuations that tend to diminish $\alpha$. This is supported by $\alpha$ measured for the ``no {\it trans}'' model, where $\xi = 0.001$. Excluding the fluctuations arising from the {\it trans} side again increases $\alpha$ for moderate $f_p$. Again, $\alpha$ for this and the unmodified model, where $\xi = 0.001$, are equal when $f_p$ is large, in which case the contribution from fluctuations are insignificant. $\alpha$ for the complete model is constant and smaller than the asymptotic value for different $f_p$. However, when the {\it trans} side is excluded, $\alpha$ diminishes with increasing $f_p$. 

In conclusion, it can be stated that inclusion of all the fluctuations present in the system is essential in order to obtain correct characteristics for the driven translocation. So, although the crowding itself does not have a significant effect on the obtained scaling, inclusion of the fluctuations pertinent to the {\it trans} side seems to be crucial for obtaining the correct scaling.

\begin{table}[ht]
\caption{Scaling exponents for translocation. Errors are $\le \pm 0.01$.} 
 \centering 
\begin{tabular}{|C{0.9cm} |C{1.4cm} |C{1.4cm}|C{1.4cm}|C{1.4cm}|C{1.4cm}|C{1.4cm}|} 
\hline\hline 
$f_p$ & $\alpha$ unmodified & $\alpha$ no \textit{trans} & $\alpha$ no \textit{cis} & $\alpha$ no pore friction & $\alpha$ no pore friction and no \textit{trans}\\ [0.5ex] 
\hline 
0.5 & 1.36 & 1.39 & 1.06 & 1.52 & 1.58\\ \hline 
1 & 1.38 & 1.40 & 1.05 & 1.53 & 1.58\\ \hline
5 & 1.40 & 1.42 & 1.05 & 1.53 & 1.55\\ \hline
10 & 1.41 & 1.42 & 1.05 & 1.52 & 1.53\\ \hline
20 & 1.43 & 1.43 & 1.03 & 1.51 & 1.51\\ \hline
40 & 1.44 & 1.43 & 1.01 & x & x\\ 
\hline 
\end{tabular}
\label{tab:ScalingExponents} 
\end{table}

\subsection{Dragged beads and waiting times}
\label{wt}

\begin{figure}[t]
\includegraphics[width=0.90\linewidth]{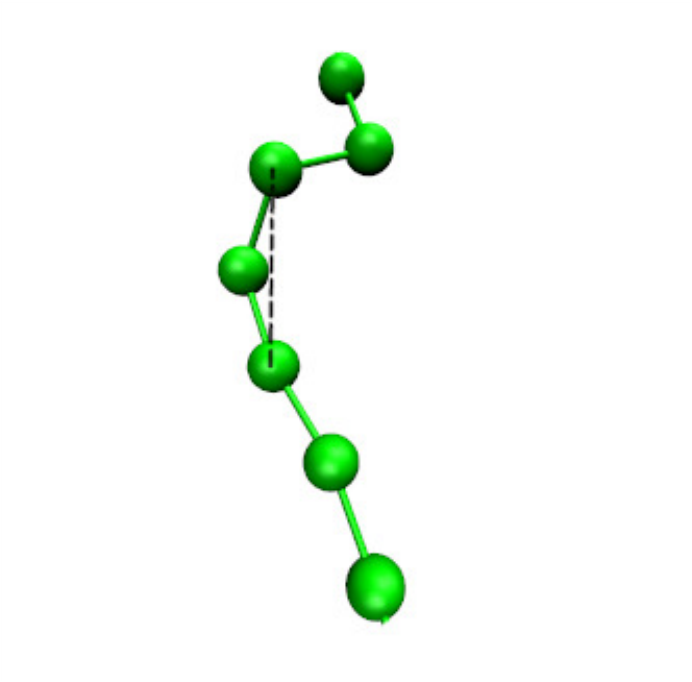}
\caption{(Color online) The distance between all polymer beads separated by two bonds, shown by the black dashed line, was measured as a function of translocation coordinate $s$.}
\label{fig:2BondLengthIllustration}
\end{figure}

We pursue further in our attempt to asses in what detail a model has to reproduce the simulated results in order to be regarded as the correct model. In what follows, we investigate in detail the relation between tension spreading and monomers on which drag is exerted. To pin down the tension propagation along the polymer during a simulation, we measure distances $l(n)$ of all bead pairs $(n-1,n+1)$ in the polymer chain that are separated by the bead $n$, see Fig.~\ref{fig:2BondLengthIllustration}. As the tension propagates from the pore on to the {\it cis} side, the polymer straightens and $l(n)$ in the tensile sections grow. This way of measuring the tension propagation is much more precise than our previous measurement based on identifying monomers moving toward the pore on the {\it cis} side~\cite{Lehtola09}.

\begin{figure}[t]
\includegraphics[width=0.90\linewidth]{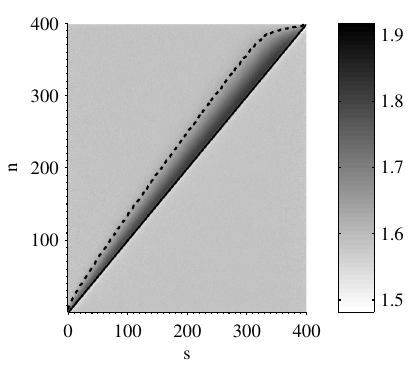}
\caption{The distribution of distances $l(n)$ of polymer beads $n-1$ and $n+1$ as a function of the translocation coordinate $s$. The shade of the pixels in the figure give the distance according to the color bar on the right.}
\label{fig:2BondLength200beads}
\end{figure}

Fig.~\ref{fig:2BondLength200beads} shows $l(n)$ as a function of the translocation coordinate $s$, {\it i.e.} the number of translocated beads, for $f_p=1$ and $N = 400$. The form of the measured distributions of $l$ with $s$ for $N = 50$, $100$, and $200$ are similar (not shown). The diagonal from the bottom left to the top right corner corresponds to the translocation coordinate. The area below the diagonal $n = s$ depicts $l(n)$ for beads translocated to the {\it trans} side and the area above it for beads on the \textit{cis} side. The qualitative picture extracted from Fig.~\ref{fig:2BondLength200beads} is what should be expected. $l(n)$ are seen to be greatest in the segments inside and immediately behind the pore on the \textit{cis} side. The segment under tension increases steadily as the beads translocate to the \textit{trans} side until it encompasses the whole polymer segment on the {\it cis} side, which then diminishes until the whole polymer chain has translocated.

\begin{figure}[t]
\includegraphics[width=0.90\linewidth]{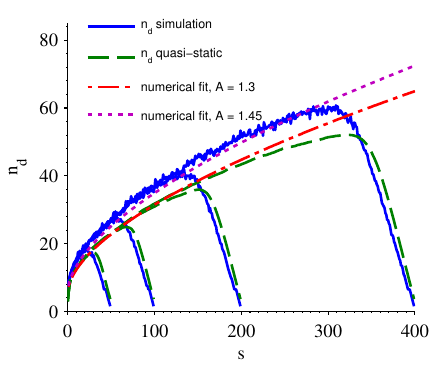}
\caption{(Color online) Number of beads in drag $n_d$ as a function of the translocation coordinate $s$ for $N = 50$, $100$, $200$, and $400$.}
\label{fig:DraggedBeadsCombined}
\end{figure}

In order to make comparison between drag force and waiting time profile, we extract a profile for the number of beads in drag from Fig.~\ref{fig:2BondLength200beads}. We take the outer contour (dashed line) of the area where the tension has propagated. Since the bond number $n$ of the vertical axis is the bead around which the distance has been calculated, we can determine the length of the tensed polymer segment in numbers of beads in drag $n_d$ by taking the vertical distance between the dashed contour and the diagonal $n = s$ for all $s$. Fig.~\ref{fig:DraggedBeadsCombined} shows lengths of the tense polymer segments calculated this way for $N = 50$, $100$, $200$, and $400$. The profiles for different $N$ are seen to be nearly identical until $n_d$ equals the number of beads still on the {\it cis} side.

\begin{figure}[t]
\includegraphics[width=0.90\linewidth]{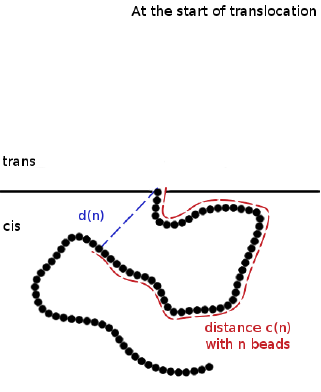}
\caption{(Color online) The quasi-static model in an initial conformation.}
\label{fig:PicForRefe1}
\end{figure}

\begin{figure}[t]
\includegraphics[width=0.90\linewidth]{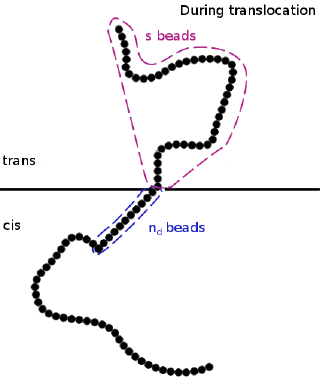}
\caption{(Color online) The quasi-static model during translocation.}
\label{fig:PicForRefe2}
\end{figure}

We compare the profile for $n_d$ to the profile resulting from a hypothetical situation where beads are pulled so rapidly that the polymer conformation has no time to respond to the pulling. In other words, the translocation velocity is much greater than the speed at which the polymer relaxes toward thermal equilibrium. Dynamics, in particular inertia, has been eliminated from this model, which we accordingly call the quasi-static model. The symbols pertinent to this model are depicted in Figs.~\ref{fig:PicForRefe1} and \ref{fig:PicForRefe2}. For clarity, the model and the resulting equations are presented without the pore beads. In the numerical fits the number of pore beads, which is a mere additive constant, is included. For this model we derive the number of beads belonging to the tense segment as a function of translocation coordinate in the following way. From each initial equilibrium conformation of our simulations we calculate the shortest distance $d(n)$ between every polymer bead $n$ and the entrance of the pore in units of average bond lengths $b$. $n_d = d(n)/b$ will be the number of beads in drag when tension reaches the bead $n$. $c(n) = n b$ is the distance from the pore entrance to the bead $n$ along the contour of the polymer. Since $n = c(n)/b$ is the number of beads originally connecting the bead $n$ to the pore entrance, $n - n_d$  is the number of beads that have already entered the pore at the time when tension reaches the bead $n$. Accordingly, the translocation coordinate is obtained as $s = n - n_d$.

When the tension just reaches the bead $n$ in the quasi-static model, the length of the tensed segment $d(n) = n_d b$ equals the initial equilibrium distance $h(n)$ from the pore entrance to the bead $n$: $h(n) = A b n^\nu = A b (s + n_d)^\nu$, where $A$ is a constant coefficient. Hence, 
\begin{equation}
n_d = A  (s + n_d)^\nu.
\label{nd}
\end{equation}

We measure $R_g$ for the initial polymer conformations to obtain $\nu \approx 0.6$. The coefficient $A$ can now be determined by solving Eq.~(\ref{nd}) for $\nu = \log(n_d/A)/\log(s + n_d)$ and requiring that $\nu \approx 0.6$ for as large range of $s$ as possible. (Close to $s = 0$ and $s = N$ Eq.~(\ref{nd}) does not hold.) We obtain $A = 1.3$. Fig.~\ref{fig:DraggedBeadsCombined} shows the length of the tense segment in numbers of beads in drag $n_d$ as a function of $s$ extracted from simulations and for the quasi-static model described above. The numerical solution of Eq~(\ref{nd}) for asymptotically long polymers follows $n_d(s)$ for the quasi-static model only to deviate from it as $s \to N$ due to retraction of the polymer tail. Our simplified model describes only the tension spreading and not the tail retraction, which could be described with $n_d$ decreasing linearly with $s$.

We can follow the same procedure to fit the numerical solution of Eq~(\ref{nd}) to $n_d$ as a function of $s$ obtained from real simulations. We  obtain $A = 1.45$. The coefficient could be made equal to that obtained for the quasi-static model by choosing a different constant tension value for extracting the profile in Fig.~\ref{fig:DraggedBeadsCombined} from Fig.~\ref{fig:2BondLength200beads}. Making the coefficients equal means making the ``$n_d$ simulation'' curve to align as closely as possible with the ``$n_d$ quasi-static'' curve. So, the deviations of ``$n_d$ quasi-static'' and ``$n_d$ simulation'' with the corresponding numerical solutions of Eq.~(\ref{nd}) are directly comparable. Hence, we see that deviation of $n_d(s)$ extracted from simulations from those for the simple quasi-static model is very small. From the $n_d$ and the numerical solutions of Eq.~(\ref{nd}) in Fig.~\ref{fig:DraggedBeadsCombined} deviation of $\alpha$ obtained for the $n_d$ from simulations and for the quasi-static model from the asymptotic value $1+\nu$ is seen to be due only to the finite polymer length.

\begin{figure}[t]
\includegraphics[width=0.90\linewidth]{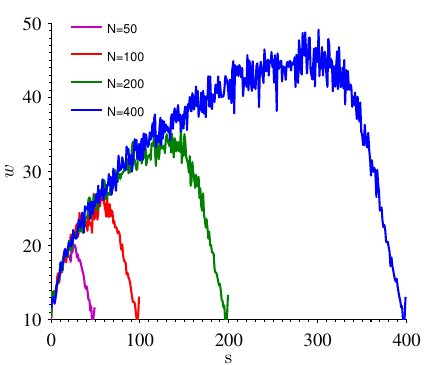}
\caption{(Color online) Waiting times $w(s)$ for $N = 50$, $100$, $200$, and $400$. The minimum $w \approx 10$ is related to how the first bead is initially placed in the pore. The final rise is due to the the total pore force decreasing with the number of beads in the pore at the end.}
\label{fig:WaitTime}
\end{figure}

Assuming $n_d$ to be directly proportional to the waiting time (for a rationalization, see below) we can use the initial conformations of our simulations and extract the waiting time profile for the quasi-static model. The real waiting time distribution gives the time that it takes (on average) for the bead $s$ to enter the \textit{trans} side after the bead $s-1$ has entered.
The actual waiting time distributions obtained from simulations for the pore force $f_p=1$ and polymers of lengths $N = 50$, $100$, $200$, and $400$ are given in Fig.~\ref{fig:WaitTime}. The forms of the waiting time and $n_d$ profiles are very similar, as is expected based on the force balance that holds for the driving and drag force. A simple force balance assumption and Langevin equation would give the velocity of the polymer in the pore as inversely proportional to the number dragged beads~\cite{Lehtola09}. This simply means that the waiting times should be proportional to the number of dragged beads or the length of the tense segment of the polymer.

\begin{figure}[t]
\includegraphics[width=0.90\linewidth]{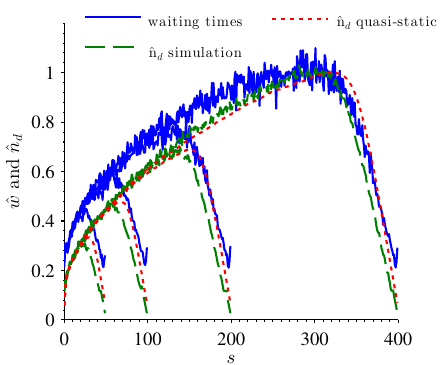}
\caption{(Color online) Scaled waiting time $\hat{w}(s)$ and number of beads in drag $\hat{n}_d(s)$ from the simulations and for the quasi-static model. Scaling was made so that the peak values of the polymers of $N = 400$ approximately match.}
\label{fig:DraggedBeadsWaitingTimeCombined}
\end{figure}

It is remarkable that based on this oversimplification a satisfactory alignment of the model and simulations can be reached, as can be seen in Fig.~\ref{fig:DraggedBeadsWaitingTimeCombined}, where $n_d$ and waiting time profiles are compared. The $n_d$ profile extracted from simulations is seen to be more rounded than the $n_d$ profile for the quasi-static model. This is due to all correlations being excluded from the quasi-static model. In the first case the boundary between monomers at rest and those set in motion is broader. In the waiting time profile correlations between monomers show most clearly, and the boundary is broadest due to the acceleration of monomers broadening the boundary between the tensed and relaxed segments and making the profile rounder.

If we integrate $n_d$ profiles to get $\tau$ for different $N$, we get $\alpha = 1.566$, $1.547$, and $1.545$ for $f_p = 1$, $5$, and $10$, respectively. From the $n_d$ profile for the quasi-static model we would get $\alpha = 1.511$. So, $\alpha$ decreases with increasing $f_p$, when only the number of monomers in drag as a function of $s$ is used instead of actual waiting times. In other words, $\alpha$ is seen to decrease when fluctuations and some aspects of the dynamics are excluded. The latter is due to ignoring that monomers that are set in motion do not reach their terminal velocity instantly but are accelerated, instead.

\section{Conclusion}\label{sec:con}

We have by Langevin dynamics simulations and varied models determined important characteristics of driven polymer translocation for finite polymer lengths $N$. Our motivation was to determine the minimum ingredients required of a driven polymer translocation model to reproduce the correct characteristics. Specifically, we addressed  how strongly the monomer crowding on the {\it trans} side modifies the overall characteristics of this process. We were also interested in the possible effect of fluctuations that we have previously found to be of importance for realistic pore force $f_p$ magnitudes~\cite{Linna12}. 

We first confirmed our previous finding for the complete model that the {\it cis} and {\it trans} sides are driven strongly out of equilibrium for realistic $f_p$. Secondly, we confirmed that the exponent $\alpha$ describing how the translocation time scales with the polymer length $\tau \sim N^{\alpha}$ increases with $f_p$. This is in accord with our previous findings~\cite{Lehtola09} and those by Dubbeldam et al.~\cite{Dubbeldam12}. The numerical model reported in~\cite{Ikonen12} based on the tension propagation on the {\it cis} side~\cite{Sakaue07} gives the opposite characteristics. Since this model completely ignores the {\it trans} side, we set out to determine, if the out-of-equilibrium dynamics of the polymer segment on the {\it trans} side is responsible for this discrepancy. This we did via modified translocation models. We found that $\alpha$ increases with $f_p$ also for the model where monomers were removed after they had translocated to the {\it trans} side. The increase was only slightly more moderate than when the {\it trans} side was included.

Setting the pore friction very close to zero resulted in $\alpha$ staying constant for varying $f_p$. This is in keeping with the expected asymptotic scaling. Only the value $\alpha$ was smaller than the asymptotic value $1 + \nu$. Excluding the {\it trans} side $\alpha$ decreased slightly with $f_p$ in the absence of pore friction but excluding the {\it cis} side we could see that the {\it trans} side has a very weak effect on the obtained scaling.

Hence, although the monomer crowding on the {\it trans} side was found to affect the scaling, its contribution was very weak. All our results on the complete and modified models could be explained by the finding of Dubbeldam et al.~\cite{Dubbeldam13} that fluctuations facilitate translocation for moderate $f_p$. Accordingly, the most drastic deficiency when excluding the {\it trans} side appears to come from eliminating the pertinent fluctuations. The same applies for the {\it cis} side. Together with the results obtained for our other models, results for the close to zero pore friction support the finding that fluctuations contribute to $\alpha$ remaining below the asymptotic value~\cite{Dubbeldam13}. In accord with our previous findings~\cite{Lehtola09,Lehtola08} the pore friction diminishes $\alpha$, which has a close connection with the finite-size effect, as analyzed in~\cite{Ikonen13}.

In order to pin down how detailed alignment with the true model is required of the model that is claimed to describe the driven polymer translocation, we also investigated in detail the tension propagation dynamics on the {\it cis} side. We measured with high precision how tension propagates along the polymer contour on the {\it cis} side in our complete dynamical model. Additionally, we produced a waiting time profile corresponding to a force-balance based approximation where the waiting time $w$ is directly proportional to the number of monomers in drag $n_d$. For this model where some dynamical aspects are ignored we find that $\alpha$ very slightly decreases with $f_p$ and approaches the value $\alpha \approx 1.55$. We then excluded all dynamics, most notably inertia, from our quasi-static model, where we used initial polymer conformations on the {\it cis} side and described the driven translocation by just applying the natural constraints for the tensed segment length and the distance of monomers from the pore in the initial equilibrium polymer conformation. For this model we obtain $\alpha \approx 1.51$. We showed that the waiting times for these models are closely reminiscent of the true waiting times obtained for the full dynamical model.

In conclusion, we found that the waiting time profiles for all the different models giving different $\alpha$ are very similar. Hence, a highly detailed alignment of the waiting times obtained for a minimal and the complete dynamical model is required to claim perfect description of the process. The increase of the scaling exponent $\alpha$ with the pore force $f_p$ for finite polymer length $N$ has now been confirmed with both Langevin dynamics and stochastic rotation dynamics. We found that for a model to reproduce correct characteristics inclusion of both the {\it cis} and the {\it trans} side with the pertinent fluctuations is necessary.

\begin{acknowledgments}
The computational resources of CSC-IT Centre for Science, Finland, and Aalto Science-IT project are acknowledged. R.P.L. thanks V.G. Rostiashvili for useful communication.
\end{acknowledgments}

\bibliography{references.bib}

\begin{thebibliography}{28}%
\makeatletter
\providecommand \@ifxundefined [1]{%
 \@ifx{#1\undefined}
}%
\providecommand \@ifnum [1]{%
 \ifnum #1\expandafter \@firstoftwo
 \else \expandafter \@secondoftwo
 \fi
}%
\providecommand \@ifx [1]{%
 \ifx #1\expandafter \@firstoftwo
 \else \expandafter \@secondoftwo
 \fi
}%
\providecommand \natexlab [1]{#1}%
\providecommand \enquote  [1]{``#1''}%
\providecommand \bibnamefont  [1]{#1}%
\providecommand \bibfnamefont [1]{#1}%
\providecommand \citenamefont [1]{#1}%
\providecommand \href@noop [0]{\@secondoftwo}%
\providecommand \href [0]{\begingroup \@sanitize@url \@href}%
\providecommand \@href[1]{\@@startlink{#1}\@@href}%
\providecommand \@@href[1]{\endgroup#1\@@endlink}%
\providecommand \@sanitize@url [0]{\catcode `\\12\catcode `\$12\catcode
  `\&12\catcode `\#12\catcode `\^12\catcode `\_12\catcode `\%12\relax}%
\providecommand \@@startlink[1]{}%
\providecommand \@@endlink[0]{}%
\providecommand \url  [0]{\begingroup\@sanitize@url \@url }%
\providecommand \@url [1]{\endgroup\@href {#1}{\urlprefix }}%
\providecommand \urlprefix  [0]{URL }%
\providecommand \Eprint [0]{\href }%
\providecommand \doibase [0]{http://dx.doi.org/}%
\providecommand \selectlanguage [0]{\@gobble}%
\providecommand \bibinfo  [0]{\@secondoftwo}%
\providecommand \bibfield  [0]{\@secondoftwo}%
\providecommand \translation [1]{[#1]}%
\providecommand \BibitemOpen [0]{}%
\providecommand \bibitemStop [0]{}%
\providecommand \bibitemNoStop [0]{.\EOS\space}%
\providecommand \EOS [0]{\spacefactor3000\relax}%
\providecommand \BibitemShut  [1]{\csname bibitem#1\endcsname}%
\let\auto@bib@innerbib\@empty
\bibitem [{\citenamefont {Kasianowicz}\ \emph {et~al.}(1996)\citenamefont
  {Kasianowicz}, \citenamefont {Brandin}, \citenamefont {Branton},\ and\
  \citenamefont {Deamer}}]{Kasianowicz96}%
  \BibitemOpen
  \bibfield  {author} {\bibinfo {author} {\bibfnamefont {J.}~\bibnamefont
  {Kasianowicz}}, \bibinfo {author} {\bibfnamefont {E.}~\bibnamefont
  {Brandin}}, \bibinfo {author} {\bibfnamefont {D.}~\bibnamefont {Branton}}, \
  and\ \bibinfo {author} {\bibfnamefont {D.}~\bibnamefont {Deamer}},\ }\href
  {http://www.pnas.org/content/93/24/13770.abstract} {\bibfield  {journal}
  {\bibinfo  {journal} {Proceedings of the National Academy of Sciences}\
  }\textbf {\bibinfo {volume} {93}},\ \bibinfo {pages} {13770} (\bibinfo {year}
  {1996})},\ \Eprint
  {http://arxiv.org/abs/http://www.pnas.org/content/93/24/13770.full.pdf+html}
  {http://www.pnas.org/content/93/24/13770.full.pdf+html} \BibitemShut
  {NoStop}%
\bibitem [{\citenamefont {Branton}\ \emph {et~al.}(2008)\citenamefont
  {Branton}, \citenamefont {Deamer}, \citenamefont {Marziali}, \citenamefont
  {Bayley}, \citenamefont {Benner}, \citenamefont {Butler}, \citenamefont
  {Di~Ventra}, \citenamefont {Garaj}, \citenamefont {Hibbs}, \citenamefont
  {Huang}, \citenamefont {Jovanovich}, \citenamefont {Krstic}, \citenamefont
  {Lindsay}, \citenamefont {Ling}, \citenamefont {Mastrangelo}, \citenamefont
  {Meller}, \citenamefont {Oliver}, \citenamefont {Pershin}, \citenamefont
  {Ramsey}, \citenamefont {Riehn}, \citenamefont {Soni}, \citenamefont
  {Tabard-Cossa}, \citenamefont {Wanunu}, \citenamefont {Wiggin},\ and\
  \citenamefont {Schloss}}]{Branton08}%
  \BibitemOpen
  \bibfield  {author} {\bibinfo {author} {\bibfnamefont {D.}~\bibnamefont
  {Branton}}, \bibinfo {author} {\bibfnamefont {D.~W.}\ \bibnamefont {Deamer}},
  \bibinfo {author} {\bibfnamefont {A.}~\bibnamefont {Marziali}}, \bibinfo
  {author} {\bibfnamefont {H.}~\bibnamefont {Bayley}}, \bibinfo {author}
  {\bibfnamefont {S.~A.}\ \bibnamefont {Benner}}, \bibinfo {author}
  {\bibfnamefont {T.}~\bibnamefont {Butler}}, \bibinfo {author} {\bibfnamefont
  {M.}~\bibnamefont {Di~Ventra}}, \bibinfo {author} {\bibfnamefont
  {S.}~\bibnamefont {Garaj}}, \bibinfo {author} {\bibfnamefont
  {A.}~\bibnamefont {Hibbs}}, \bibinfo {author} {\bibfnamefont
  {X.}~\bibnamefont {Huang}}, \bibinfo {author} {\bibfnamefont {S.~B.}\
  \bibnamefont {Jovanovich}}, \bibinfo {author} {\bibfnamefont {P.~S.}\
  \bibnamefont {Krstic}}, \bibinfo {author} {\bibfnamefont {S.}~\bibnamefont
  {Lindsay}}, \bibinfo {author} {\bibfnamefont {X.~S.}\ \bibnamefont {Ling}},
  \bibinfo {author} {\bibfnamefont {C.~H.}\ \bibnamefont {Mastrangelo}},
  \bibinfo {author} {\bibfnamefont {A.}~\bibnamefont {Meller}}, \bibinfo
  {author} {\bibfnamefont {J.~S.}\ \bibnamefont {Oliver}}, \bibinfo {author}
  {\bibfnamefont {Y.~V.}\ \bibnamefont {Pershin}}, \bibinfo {author}
  {\bibfnamefont {J.~M.}\ \bibnamefont {Ramsey}}, \bibinfo {author}
  {\bibfnamefont {R.}~\bibnamefont {Riehn}}, \bibinfo {author} {\bibfnamefont
  {G.~V.}\ \bibnamefont {Soni}}, \bibinfo {author} {\bibfnamefont
  {V.}~\bibnamefont {Tabard-Cossa}}, \bibinfo {author} {\bibfnamefont
  {M.}~\bibnamefont {Wanunu}}, \bibinfo {author} {\bibfnamefont
  {M.}~\bibnamefont {Wiggin}}, \ and\ \bibinfo {author} {\bibfnamefont {J.~A.}\
  \bibnamefont {Schloss}},\ }\href {\doibase 10.1038/nbt.1495} {\bibfield
  {journal} {\bibinfo  {journal} {Nature Biotechnology}\ }\textbf {\bibinfo
  {volume} {26}},\ \bibinfo {pages} {1146} (\bibinfo {year}
  {2008})}\BibitemShut {NoStop}%
\bibitem [{\citenamefont {Wanunu}(2012)}]{Wanunu12}%
  \BibitemOpen
  \bibfield  {author} {\bibinfo {author} {\bibfnamefont {M.}~\bibnamefont
  {Wanunu}},\ }\href {\doibase http://dx.doi.org/10.1016/j.plrev.2012.05.010}
  {\bibfield  {journal} {\bibinfo  {journal} {Physics of Life Reviews}\
  }\textbf {\bibinfo {volume} {9}},\ \bibinfo {pages} {125 } (\bibinfo {year}
  {2012})}\BibitemShut {NoStop}%
\bibitem [{\citenamefont {Alberts}\ \emph {et~al.}(1994)\citenamefont
  {Alberts}, \citenamefont {Bray}, \citenamefont {Lewis}, \citenamefont {Raff},
  \citenamefont {Roberts},\ and\ \citenamefont {Watson}}]{Alberts94}%
  \BibitemOpen
  \bibfield  {author} {\bibinfo {author} {\bibfnamefont {B.}~\bibnamefont
  {Alberts}}, \bibinfo {author} {\bibfnamefont {D.}~\bibnamefont {Bray}},
  \bibinfo {author} {\bibfnamefont {J.}~\bibnamefont {Lewis}}, \bibinfo
  {author} {\bibfnamefont {M.}~\bibnamefont {Raff}}, \bibinfo {author}
  {\bibfnamefont {K.}~\bibnamefont {Roberts}}, \ and\ \bibinfo {author}
  {\bibfnamefont {J.~D.}\ \bibnamefont {Watson}},\ }\href
  {http://www.amazon.com/exec/obidos/redirect?tag=citeulike07-20\&path=ASIN/0815316194}
  {\emph {\bibinfo {title} {{Molecular Biology of the Cell}}}},\ \bibinfo
  {edition} {3rd}\ ed.\ (\bibinfo  {publisher} {Garland Science},\ \bibinfo
  {year} {1994})\BibitemShut {NoStop}%
\bibitem [{\citenamefont {Sung}\ and\ \citenamefont {Park}(1996)}]{Sung96}%
  \BibitemOpen
  \bibfield  {author} {\bibinfo {author} {\bibfnamefont {W.}~\bibnamefont
  {Sung}}\ and\ \bibinfo {author} {\bibfnamefont {P.~J.}\ \bibnamefont
  {Park}},\ }\href@noop {} {\bibfield  {journal} {\bibinfo  {journal} {Phys.
  Rev. Lett.}\ }\textbf {\bibinfo {volume} {77}},\ \bibinfo {pages} {783}
  (\bibinfo {year} {1996})}\BibitemShut {NoStop}%
\bibitem [{\citenamefont {Muthukumar}(1999)}]{Muthukumar99}%
  \BibitemOpen
  \bibfield  {author} {\bibinfo {author} {\bibfnamefont {M.}~\bibnamefont
  {Muthukumar}},\ }\href {\doibase 10.1063/1.480386} {\bibfield  {journal}
  {\bibinfo  {journal} {The Journal of Chemical Physics}\ }\textbf {\bibinfo
  {volume} {111}},\ \bibinfo {pages} {10371} (\bibinfo {year}
  {1999})}\BibitemShut {NoStop}%
\bibitem [{\citenamefont {Chuang}\ \emph {et~al.}(2001)\citenamefont {Chuang},
  \citenamefont {Kantor},\ and\ \citenamefont {Kardar}}]{Chuang01}%
  \BibitemOpen
  \bibfield  {author} {\bibinfo {author} {\bibfnamefont {J.}~\bibnamefont
  {Chuang}}, \bibinfo {author} {\bibfnamefont {Y.}~\bibnamefont {Kantor}}, \
  and\ \bibinfo {author} {\bibfnamefont {M.}~\bibnamefont {Kardar}},\
  }\href@noop {} {\bibfield  {journal} {\bibinfo  {journal} {Phys. Rev. E}\
  }\textbf {\bibinfo {volume} {65}},\ \bibinfo {pages} {011802} (\bibinfo
  {year} {2001})}\BibitemShut {NoStop}%
\bibitem [{\citenamefont {Kantor}\ and\ \citenamefont
  {Kardar}(2004)}]{Kantor04}%
  \BibitemOpen
  \bibfield  {author} {\bibinfo {author} {\bibfnamefont {Y.}~\bibnamefont
  {Kantor}}\ and\ \bibinfo {author} {\bibfnamefont {M.}~\bibnamefont
  {Kardar}},\ }\href {\doibase 10.1103/PhysRevE.69.021806} {\bibfield
  {journal} {\bibinfo  {journal} {Phys. Rev. E}\ }\textbf {\bibinfo {volume}
  {69}},\ \bibinfo {pages} {021806} (\bibinfo {year} {2004})}\BibitemShut
  {NoStop}%
\bibitem [{\citenamefont {Lehtola}\ \emph {et~al.}(2009)\citenamefont
  {Lehtola}, \citenamefont {Linna},\ and\ \citenamefont {Kaski}}]{Lehtola09}%
  \BibitemOpen
  \bibfield  {author} {\bibinfo {author} {\bibfnamefont {V.~V.}\ \bibnamefont
  {Lehtola}}, \bibinfo {author} {\bibfnamefont {R.~P.}\ \bibnamefont {Linna}},
  \ and\ \bibinfo {author} {\bibfnamefont {K.}~\bibnamefont {Kaski}},\ }\href
  {http://stacks.iop.org/0295-5075/85/i=5/a=58006} {\bibfield  {journal}
  {\bibinfo  {journal} {EPL (Europhysics Letters)}\ }\textbf {\bibinfo {volume}
  {85}},\ \bibinfo {pages} {58006} (\bibinfo {year} {2009})}\BibitemShut
  {NoStop}%
\bibitem [{\citenamefont {Lehtola}\ \emph {et~al.}(2008)\citenamefont
  {Lehtola}, \citenamefont {Linna},\ and\ \citenamefont {Kaski}}]{Lehtola08}%
  \BibitemOpen
  \bibfield  {author} {\bibinfo {author} {\bibfnamefont {V.~V.}\ \bibnamefont
  {Lehtola}}, \bibinfo {author} {\bibfnamefont {R.~P.}\ \bibnamefont {Linna}},
  \ and\ \bibinfo {author} {\bibfnamefont {K.}~\bibnamefont {Kaski}},\ }\href
  {\doibase 10.1103/PhysRevE.78.061803} {\bibfield  {journal} {\bibinfo
  {journal} {Phys. Rev. E}\ }\textbf {\bibinfo {volume} {78}},\ \bibinfo
  {pages} {061803} (\bibinfo {year} {2008})}\BibitemShut {NoStop}%
\bibitem [{\citenamefont {Sakaue}(2007)}]{Sakaue07}%
  \BibitemOpen
  \bibfield  {author} {\bibinfo {author} {\bibfnamefont {T.}~\bibnamefont
  {Sakaue}},\ }\href {\doibase 10.1103/PhysRevE.76.021803} {\bibfield
  {journal} {\bibinfo  {journal} {Phys. Rev. E}\ }\textbf {\bibinfo {volume}
  {76}},\ \bibinfo {pages} {021803} (\bibinfo {year} {2007})}\BibitemShut
  {NoStop}%
\bibitem [{\citenamefont {Dubbeldam}\ \emph {et~al.}(2012)\citenamefont
  {Dubbeldam}, \citenamefont {Rostiashvili}, \citenamefont {Milchev},\ and\
  \citenamefont {Vilgis}}]{Dubbeldam12}%
  \BibitemOpen
  \bibfield  {author} {\bibinfo {author} {\bibfnamefont {J.~L.~A.}\
  \bibnamefont {Dubbeldam}}, \bibinfo {author} {\bibfnamefont {V.~G.}\
  \bibnamefont {Rostiashvili}}, \bibinfo {author} {\bibfnamefont
  {A.}~\bibnamefont {Milchev}}, \ and\ \bibinfo {author} {\bibfnamefont
  {T.~A.}\ \bibnamefont {Vilgis}},\ }\href {\doibase
  10.1103/PhysRevE.85.041801} {\bibfield  {journal} {\bibinfo  {journal} {Phys.
  Rev. E}\ }\textbf {\bibinfo {volume} {85}},\ \bibinfo {pages} {041801}
  (\bibinfo {year} {2012})}\BibitemShut {NoStop}%
\bibitem [{\citenamefont {Ikonen}\ \emph {et~al.}(2012)\citenamefont {Ikonen},
  \citenamefont {Bhattacharya}, \citenamefont {Ala-Nissila},\ and\
  \citenamefont {Sung}}]{Ikonen12}%
  \BibitemOpen
  \bibfield  {author} {\bibinfo {author} {\bibfnamefont {T.}~\bibnamefont
  {Ikonen}}, \bibinfo {author} {\bibfnamefont {A.}~\bibnamefont
  {Bhattacharya}}, \bibinfo {author} {\bibfnamefont {T.}~\bibnamefont
  {Ala-Nissila}}, \ and\ \bibinfo {author} {\bibfnamefont {W.}~\bibnamefont
  {Sung}},\ }\href@noop {} {\bibfield  {journal} {\bibinfo  {journal} {Phys.
  Rev. E}\ }\textbf {\bibinfo {volume} {85}},\ \bibinfo {pages} {051803}
  (\bibinfo {year} {2012})}\BibitemShut {NoStop}%
\bibitem [{\citenamefont {Saito}\ and\ \citenamefont {Sakaue}(2012)}]{Saito12}%
  \BibitemOpen
  \bibfield  {author} {\bibinfo {author} {\bibfnamefont {T.}~\bibnamefont
  {Saito}}\ and\ \bibinfo {author} {\bibfnamefont {T.}~\bibnamefont {Sakaue}},\
  }\href@noop {} {\bibfield  {journal} {\bibinfo  {journal} {Phys. Rev. E}\
  }\textbf {\bibinfo {volume} {85}},\ \bibinfo {pages} {061803} (\bibinfo
  {year} {2012})}\BibitemShut {NoStop}%
\bibitem [{\citenamefont {Rowghanian}\ and\ \citenamefont
  {Grosberg}(2011)}]{Rowghanian11}%
  \BibitemOpen
  \bibfield  {author} {\bibinfo {author} {\bibfnamefont {P.}~\bibnamefont
  {Rowghanian}}\ and\ \bibinfo {author} {\bibfnamefont {A.}~\bibnamefont
  {Grosberg}},\ }\href@noop {} {\bibfield  {journal} {\bibinfo  {journal} {J.
  Phys. Chem. B}\ }\textbf {\bibinfo {volume} {115}},\ \bibinfo {pages} {14127}
  (\bibinfo {year} {2011})}\BibitemShut {NoStop}%
\bibitem [{\citenamefont {Ikonen}\ \emph {et~al.}(2013)\citenamefont {Ikonen},
  \citenamefont {Bhattacharya}, \citenamefont {Ala-Nissila},\ and\
  \citenamefont {Sung}}]{Ikonen13}%
  \BibitemOpen
  \bibfield  {author} {\bibinfo {author} {\bibfnamefont {T.}~\bibnamefont
  {Ikonen}}, \bibinfo {author} {\bibfnamefont {A.}~\bibnamefont
  {Bhattacharya}}, \bibinfo {author} {\bibfnamefont {T.}~\bibnamefont
  {Ala-Nissila}}, \ and\ \bibinfo {author} {\bibfnamefont {W.}~\bibnamefont
  {Sung}},\ }\href@noop {} {\bibfield  {journal} {\bibinfo  {journal} {EPL}\
  }\textbf {\bibinfo {volume} {103}},\ \bibinfo {pages} {38001} (\bibinfo
  {year} {2013})}\BibitemShut {NoStop}%
\bibitem [{\citenamefont {Dubbeldam}\ \emph {et~al.}(2013)\citenamefont
  {Dubbeldam}, \citenamefont {Rostiashvili}, \citenamefont {Milchev},\ and\
  \citenamefont {Vilgis}}]{Dubbeldam13}%
  \BibitemOpen
  \bibfield  {author} {\bibinfo {author} {\bibfnamefont {J.~L.~A.}\
  \bibnamefont {Dubbeldam}}, \bibinfo {author} {\bibfnamefont {V.~G.}\
  \bibnamefont {Rostiashvili}}, \bibinfo {author} {\bibfnamefont
  {A.}~\bibnamefont {Milchev}}, \ and\ \bibinfo {author} {\bibfnamefont
  {T.~A.}\ \bibnamefont {Vilgis}},\ }\href@noop {} {\bibfield  {journal}
  {\bibinfo  {journal} {Phys. Rev. E}\ }\textbf {\bibinfo {volume} {87}},\
  \bibinfo {pages} {032147} (\bibinfo {year} {2013})}\BibitemShut {NoStop}%
\bibitem [{\citenamefont {Dubbeldam}\ \emph {et~al.}()\citenamefont
  {Dubbeldam}, \citenamefont {Rostiashvili},\ and\ \citenamefont
  {Vilgis}}]{Dubbeldam14}%
  \BibitemOpen
  \bibfield  {author} {\bibinfo {author} {\bibfnamefont {J.~L.~A.}\
  \bibnamefont {Dubbeldam}}, \bibinfo {author} {\bibfnamefont {V.~G.}\
  \bibnamefont {Rostiashvili}}, \ and\ \bibinfo {author} {\bibfnamefont
  {T.~A.}\ \bibnamefont {Vilgis}},\ }\href@noop {} {\bibinfo  {journal}
  {http://arxiv.org/pdf/1404.0167v1.pdf}\ }\BibitemShut {NoStop}%
\bibitem [{\citenamefont {Tinland}\ \emph {et~al.}(1997)\citenamefont
  {Tinland}, \citenamefont {Pluen}, \citenamefont {Sturm},\ and\ \citenamefont
  {Weill}}]{Tinland97}%
  \BibitemOpen
\bibfield  {journal} {  }\bibfield  {author} {\bibinfo {author} {\bibfnamefont
  {B.}~\bibnamefont {Tinland}}, \bibinfo {author} {\bibfnamefont
  {A.}~\bibnamefont {Pluen}}, \bibinfo {author} {\bibfnamefont
  {J.}~\bibnamefont {Sturm}}, \ and\ \bibinfo {author} {\bibfnamefont
  {G.}~\bibnamefont {Weill}},\ }\href@noop {} {\bibfield  {journal} {\bibinfo
  {journal} {Macromolecules}\ }\textbf {\bibinfo {volume} {30}},\ \bibinfo
  {pages} {5763} (\bibinfo {year} {1997})}\BibitemShut {NoStop}%
\bibitem [{\citenamefont {Ermak}\ and\ \citenamefont
  {Buckholz}(1980)}]{Ermak80}%
  \BibitemOpen
  \bibfield  {author} {\bibinfo {author} {\bibfnamefont {D.~L.}\ \bibnamefont
  {Ermak}}\ and\ \bibinfo {author} {\bibfnamefont {H.}~\bibnamefont
  {Buckholz}},\ }\href {\doibase
  http://dx.doi.org/10.1016/0021-9991(80)90084-4} {\bibfield  {journal}
  {\bibinfo  {journal} {Journal of Computational Physics}\ }\textbf {\bibinfo
  {volume} {35}},\ \bibinfo {pages} {169 } (\bibinfo {year}
  {1980})}\BibitemShut {NoStop}%
\bibitem [{\citenamefont {van Gunsteren}\ and\ \citenamefont
  {Berendsen}(1977)}]{vanGunsteren77}%
  \BibitemOpen
  \bibfield  {author} {\bibinfo {author} {\bibfnamefont {W.}~\bibnamefont {van
  Gunsteren}}\ and\ \bibinfo {author} {\bibfnamefont {H.}~\bibnamefont
  {Berendsen}},\ }\href {\doibase 10.1080/00268977700102571} {\bibfield
  {journal} {\bibinfo  {journal} {Molecular Physics}\ }\textbf {\bibinfo
  {volume} {34}},\ \bibinfo {pages} {1311} (\bibinfo {year} {1977})},\ \Eprint
  {http://arxiv.org/abs/http://www.tandfonline.com/doi/pdf/10.1080/00268977700102571}
  {http://www.tandfonline.com/doi/pdf/10.1080/00268977700102571} \BibitemShut
  {NoStop}%
\bibitem [{\citenamefont {Allen}\ and\ \citenamefont
  {Tildesley}(2006)}]{Allen}%
  \BibitemOpen
  \bibfield  {author} {\bibinfo {author} {\bibfnamefont {M.~P.}\ \bibnamefont
  {Allen}}\ and\ \bibinfo {author} {\bibfnamefont {D.~J.}\ \bibnamefont
  {Tildesley}},\ }\href@noop {} {\emph {\bibinfo {title} {Computer Simulation
  of Liquids}}}\ (\bibinfo  {publisher} {Clarendon Press},\ \bibinfo {address}
  {Oxford},\ \bibinfo {year} {2006})\BibitemShut {NoStop}%
\bibitem [{\citenamefont {Sauer-Budge}\ \emph {et~al.}(2003)\citenamefont
  {Sauer-Budge}, \citenamefont {Nyamwanda}, \citenamefont {Lubensky},\ and\
  \citenamefont {Branton}}]{Sauerbudge03}%
  \BibitemOpen
  \bibfield  {author} {\bibinfo {author} {\bibfnamefont {A.~F.}\ \bibnamefont
  {Sauer-Budge}}, \bibinfo {author} {\bibfnamefont {J.~A.}\ \bibnamefont
  {Nyamwanda}}, \bibinfo {author} {\bibfnamefont {D.~K.}\ \bibnamefont
  {Lubensky}}, \ and\ \bibinfo {author} {\bibfnamefont {D.}~\bibnamefont
  {Branton}},\ }\href {\doibase 10.1103/PhysRevLett.90.238101} {\bibfield
  {journal} {\bibinfo  {journal} {Phys. Rev. Lett.}\ }\textbf {\bibinfo
  {volume} {90}},\ \bibinfo {pages} {238101} (\bibinfo {year}
  {2003})}\BibitemShut {NoStop}%
\bibitem [{\citenamefont {Luan}\ and\ \citenamefont
  {Aksimentiev}(2008)}]{Luan08}%
  \BibitemOpen
  \bibfield  {author} {\bibinfo {author} {\bibfnamefont {B.}~\bibnamefont
  {Luan}}\ and\ \bibinfo {author} {\bibfnamefont {A.}~\bibnamefont
  {Aksimentiev}},\ }\href@noop {} {\bibfield  {journal} {\bibinfo  {journal}
  {Phys. Rev. E}\ }\textbf {\bibinfo {volume} {78}},\ \bibinfo {pages} {021912}
  (\bibinfo {year} {2008})}\BibitemShut {NoStop}%
\bibitem [{\citenamefont {Keyser}\ \emph {et~al.}(2006)\citenamefont {Keyser},
  \citenamefont {Koeleman}, \citenamefont {Dorp}, \citenamefont {Krapf},
  \citenamefont {Smeets}, \citenamefont {R.}, \citenamefont {Lemay},
  \citenamefont {Dekker},\ and\ \citenamefont {Dekker}}]{Keyser06}%
  \BibitemOpen
  \bibfield  {author} {\bibinfo {author} {\bibfnamefont {U.}~\bibnamefont
  {Keyser}}, \bibinfo {author} {\bibfnamefont {B.}~\bibnamefont {Koeleman}},
  \bibinfo {author} {\bibfnamefont {S.}~\bibnamefont {Dorp}}, \bibinfo {author}
  {\bibfnamefont {D.}~\bibnamefont {Krapf}}, \bibinfo {author} {\bibnamefont
  {Smeets}}, \bibinfo {author} {\bibnamefont {R.}}, \bibinfo {author}
  {\bibfnamefont {S.}~\bibnamefont {Lemay}}, \bibinfo {author} {\bibfnamefont
  {N.}~\bibnamefont {Dekker}}, \ and\ \bibinfo {author} {\bibfnamefont
  {C.}~\bibnamefont {Dekker}},\ }\href@noop {} {\bibfield  {journal} {\bibinfo
  {journal} {Nature Physics}\ }\textbf {\bibinfo {volume} {2}},\ \bibinfo
  {pages} {473} (\bibinfo {year} {2006})}\BibitemShut {NoStop}%
\bibitem [{\citenamefont {Linna}\ and\ \citenamefont {Kaski}(2012)}]{Linna12}%
  \BibitemOpen
  \bibfield  {author} {\bibinfo {author} {\bibfnamefont {R.~P.}\ \bibnamefont
  {Linna}}\ and\ \bibinfo {author} {\bibfnamefont {K.}~\bibnamefont {Kaski}},\
  }\href@noop {} {\bibfield  {journal} {\bibinfo  {journal} {Phys. Rev. E}\
  }\textbf {\bibinfo {volume} {85}},\ \bibinfo {pages} {041910} (\bibinfo
  {year} {2012})}\BibitemShut {NoStop}%
\bibitem [{\citenamefont {Linna}\ \emph {et~al.}(2014)\citenamefont {Linna},
  \citenamefont {Moisio}, \citenamefont {Suhonen},\ and\ \citenamefont
  {Kaski}}]{Linna14}%
  \BibitemOpen
  \bibfield  {author} {\bibinfo {author} {\bibfnamefont {R.~P.}\ \bibnamefont
  {Linna}}, \bibinfo {author} {\bibfnamefont {J.~E.}\ \bibnamefont {Moisio}},
  \bibinfo {author} {\bibfnamefont {P.~M.}\ \bibnamefont {Suhonen}}, \ and\
  \bibinfo {author} {\bibfnamefont {K.}~\bibnamefont {Kaski}},\ }\href@noop {}
  {\bibfield  {journal} {\bibinfo  {journal} {Phys. Rev. E}\ }\textbf {\bibinfo
  {volume} {89}},\ \bibinfo {pages} {052702} (\bibinfo {year}
  {2014})}\BibitemShut {NoStop}%
\bibitem [{\citenamefont {Lehtola}\ \emph {et~al.}(2010)\citenamefont
  {Lehtola}, \citenamefont {Kaski},\ and\ \citenamefont {Linna}}]{Lehtola10}%
  \BibitemOpen
  \bibfield  {author} {\bibinfo {author} {\bibfnamefont {V.~V.}\ \bibnamefont
  {Lehtola}}, \bibinfo {author} {\bibfnamefont {K.}~\bibnamefont {Kaski}}, \
  and\ \bibinfo {author} {\bibfnamefont {R.~P.}\ \bibnamefont {Linna}},\
  }\href@noop {} {\bibfield  {journal} {\bibinfo  {journal} {Phys. Rev. E}\
  }\textbf {\bibinfo {volume} {82}},\ \bibinfo {pages} {031908} (\bibinfo
  {year} {2010})}\BibitemShut {NoStop}%
\end{thebibliography}%

\end{document}